\documentclass[sigconf]{acmart}

\usepackage{graphicx}
\usepackage{todonotes}
\usepackage{listings}
\usepackage[nolist,withpage]{acronym}
\usepackage{amsmath}
\usepackage{textcomp}
\usepackage{multirow}
\usepackage{multicol}
\usepackage{changepage}
\usepackage{algorithm}		  
\usepackage{algpseudocode}	  
\usepackage{subcaption}
\usepackage{arydshln} 
\usepackage{enumitem}

\newlist{renum}{enumerate}{1}
\setlist[renum,1]{
    label=(R\arabic*),
    leftmargin=*
}

\newcommand{\etal}{~et~al.~}

\newcommand{\spc}{SPOQchain}
\newcommand{\spshort}{SPOQ}
\AtBeginDocument{%
  }

\usepackage{amssymb}

\begin{document}

\settopmatter{printfolios=true,printacmref=false}
\renewcommand\footnotetextcopyrightpermission[1]{} 
\pagestyle{plain} 


\title{\spc: Platform for Secure, Scalable, and Privacy-Preserving Supply Chain Tracing and Counterfeit Protection}

\author{Moritz Finke}
\email{moritz.finke@uni-wuerzburg.de}
\author{Alexandra Dmitrienko}
\email{alexandra.dmitrienko@uni-wuerzburg.de}
\author{Jasper Stang}
\email{jasper.stang@uni-wuerzburg.de}
\affiliation{%
  \institution{Julius-Maximilians Universität Würzburg}
  \city{Würzburg}
  \state{Bavaria}
  \country{Germany}
}

\renewcommand{\shortauthors}{Finke et al.}

\begin{abstract}
Product lifecycle tracing is increasingly in the focus of regulators and
producers, as shown with the initiative of the \ac{dpp}~\cite{jansen2023stop}.
Likewise, new methods of counterfeit detection are developed that are, e.g.,
based on \acp{puf}. In order to ensure trust and integrity of product lifecycle
data, multiple existing supply chain tracing systems are built on blockchain
technology. However, only few solutions employ secure identifiers such as
\acp{puf}. Furthermore, existing systems that publish the data of individual
products, in part fully transparently, have a detrimental impact on scalability
and the privacy of users. This work proposes \spc, a novel blockchain-based platform that provides comprehensive lifecycle traceability and originality verification while ensuring high efficiency and user privacy.
The improved efficiency is achieved by a sophisticated batching mechanism that removes lifecycle redundancies. In addition to the successful evaluation of \spc's scalability, this work provides a comprehensive analysis of privacy and security aspects, demonstrating the need and qualification of \spc~for the future of supply chain tracing.

\end{abstract}



\received{30 August 2024}

\maketitle

\section{Introduction} \label{sec:intro}

In today's globalized and highly competitive market environment, one of the most pressing challenges is the proliferation of counterfeit goods. Counterfeiting not only undermines brand integrity and customer trust but also poses a hazard to the health and safety of consumers. For instance, $97\%$~of all counterfeit non-food items detected by the European Commission between 2010 and 2017 were found to
pose a serious risk~\cite{doi/10.2814/165063}. Overall,
$3.3\%$~of total world trade is attributed to counterfeits
and pirated products, underscoring the need for effective anti-counterfeiting
methods~\cite{g2g9f533-en}. Moreover, the ability to trace product lifecycles is increasingly promoted
through initiatives such as the \acf{dpp}~\cite{jansen2023stop}. Additionally,
the capability to verify the authenticity of products is not only required by
supply chain actors, but remains relevant throughout the entire product
lifecycle. For instance, on \ac{c2c} e-commerce platforms, the unknown origin of
products currently necessitates significant manual labor for authenticity
verification~\cite{li2023asymmetric}.

\vspace{0.1cm} \noindent \textbf{Existing Solutions} Common solutions to these
demands are block-chain-based supply chain and product lifecycle
tracing~\cite{abeyratne2016blockchain,shaffan2020blockchain,helo2020real,toyoda2017novel,caro2018blockchain,baralla2019ensure,marchese2021agri,westerkamp2018blockchain,aniello2021anti,negka2019employing,kennedy2017enhanced,felicetti2023deep}.
Using blockchains circumvents the reliance on
central authorities to manage product data. Typically, a product's lifecycle
information that is logged via smart contracts includes product descriptions, physical identification data, past owner lists, and an extensible events list recorded throughout its existence~\cite{abeyratne2016blockchain}.

Generally, two types of approaches exist: Those publishing individual
products in the
blockchain~\cite{abeyratne2016blockchain,shaffan2020blockchain,helo2020real,toyoda2017novel,caro2018blockchain,aniello2021anti,negka2019employing,kennedy2017enhanced,felicetti2023deep},
and those tracing batches without acknowledging identities of individual
goods~\cite{baralla2019ensure,marchese2021agri,westerkamp2018blockchain}.
Only one solution~\cite{abeyratne2016blockchain} considers the association of a published batch with its
individually traceable products. 

\vspace{0.1cm}\noindent \textbf{Challenges} The well-known issue of
limited throughput and replicated data storage of blockchains presents significant bottlenecks for
global system deployments~\cite{vde2020spec}. As a result, solutions that publish comprehensive
product information on the
blockchain~\cite{shaffan2020blockchain,toyoda2017novel,abeyratne2016blockchain}
can be assumed to face scalability limitations. On the other hand, more efficient systems that
only trace batches do not have the capability to verify
individual goods and assess their custom lifecycles. The only solution~\cite{abeyratne2016blockchain} that supports both, tracing of batches and individual batches, still mandates registration of all individual products on the blockchain, leading to high redundancies and the risk of overwhelming the capacity of blockchains. Additionally, publication of lifecycle data on blockchains poses significant privacy concerns. For instance, this data could reveal the current location of shipments and the personal information or residence of end customers.

Only few systems~\cite{aniello2021anti,negka2019employing,kennedy2017enhanced,felicetti2023deep} consider anti-counterfeiting measures that acknowledge that simple and cost-effective identification methods such as QR codes~\cite{toyoda2017novel} are insufficient to protect against product replication. These systems incorporate sophisticated approaches such as
optical and electronic \acfp{puf}~\cite{mcgrath2019puf}. \acp{puf} map physical properties that are, e.g., inherent to a product, to digital representations that are unique and extremely challenging to replicate. However, existing \ac{puf}-based supply chain tracing solutions~\cite{aniello2021anti,negka2019employing,kennedy2017enhanced,felicetti2023deep} focus only on specific \acp{puf}, reducing the compatibility with different product types. Hence, a notable challenge faced by existing systems is the interoperability with diverse mapping methods.

\vspace{0.1cm} 
\noindent\textbf{Contributions} In this work, we propose \spc, a novel supply chain tracing and product lifecycle and anti-counterfeiting solution addressing the described efficiency, privacy, and interoperability challenges. Particularly, we make the following contributions:  

\begin{itemize}

\item \text{Design:} We propose the design of \spc~that introduces
efficient batch-based tracing with \emph{on demand product publication}. With this approach, individual product lifecycles are only recorded when it is justified, leading to greatly reduced blockchain workload.
Furthermore, \spc~ensures user privacy and allows users to retain data
ownership. To further reduce blockchain-based bottlenecks, it leverages a
general concept of external storage systems. For increased security, it provides support
for \ac{puf}-based product identification methods while maintaining
interoperability with other (e.g., more cost-effective but less clone-resilient)
methods.

\item \text{Implementation:} We implement the core functionality of \spc~as
a portable library in Go~\cite{golang} that can be embedded into existing supply
chain management applications. Our implementation targets Hyperledger Sawtooth
blockchain~\cite{sawtooth} and offers an Android-based client. To demonstrate
flexibility in terms of supported storage systems, our implementation
exemplarily supports two storage solutions, one cloud-based server that can be
self-hosted by, e.g., producers, and the other one based on the distributed
\ac{ipfs}~\cite{benet2014ipfs}.

\item \text{Evaluation:} We provide a comprehensive evaluation in regard to the efficiency improvements achieved by our novel batching mechanism in terms
of transaction volume and storage requirements. The results confirm a linear reduction for both properties. In addition, we
evaluate security and privacy guarantees by conducting an~(informal)~security
and privacy analysis.
\end{itemize}

Overall, \spc~is the first solution supporting efficient batch-based tracing while providing on demand individual product tracing. Our evaluation demonstrates its strong security and privacy guarantees, while maintaining improved transaction throughput and storage efficiency.

\vspace{0.1cm}
\noindent\textbf{Outline} The paper explores blockchains and \acp{puf} in Section~\ref{sec:background}. We elaborate on the system design in Section~\ref{sec:design}, detail its implementation in Section~\ref{sec:impl}, and evaluate its efficiency and security in Sections~\ref{sec:efficiency} and \ref{sec:eval}. Related work is reviewed in Section~\ref{sec:relwork}. Finally, the paper is concluded in Section~\ref{sec:conc}.

\section{Background}
\label{sec:background}
This section provides essential background knowledge on blockchains and \acp{puf}.

\subsection{Blockchain}
\label{sec:background:blockchain}

Supply chain tracing solutions commonly rely on blockchain, a distributed append-only ledger technology whose integrity is guarded by its operators (validators)~\cite{buterin2014next}.
Blockchains such as Ethereum~\cite{buterin2014next} and Hyperledger Sawtooth~\cite{sawtooth} keep track of all past transactions and share this trail
with a (public) network~\cite{crosby2016blockchain}.
New transactions are summarized and published as a block.
Every new block
builds on the previous one, e.g., by including its hash value~\cite{nakamoto2008bitcoin}.
The recording of all transactions and continued hashing of past states
ensures that past transactions cannot be forged without detection.
In public blockchains, any interested actor can participate in its hosting. In contrast, permissioned blockchains define the set of blockchain hosts and restrict access to certain interfaces.

Blockchains require a consensus algorithm that manages the transition to new
blockchain states. Such consensus can be \ac{bft}, allowing
a limited number of nodes to be malicious without affecting functionality and
integrity. In contrast, \ac{cft} algorithms only
tolerate a certain amount of nodes to become
unavailable.~\cite{xiao2019distributed}

Originally designed for decentralized electronic payments (cf.
Bitcoin~\cite{nakamoto2008bitcoin}), blockchains such as
Ethereum~\cite{buterin2014next} evolved to support execution of
\emph{smart contracts}. These computer programs run reliably across the
blockchain network, executed by participating nodes. A more adaptable and
feature-rich alternative to public blockchains is found with permissioned
solutions such as Hyperledger Sawtooth that also employ efficient consensus
algorithms~\cite{sawtooth}.

\subsection{\acfp{puf}}
\label{sec:background:puf}

Conventional identifiers such as barcodes lack security due to their vulnerability to replication. \acp{puf} offer an alternative, as they generate a unique response when challenged. This response is derived from inherent random electronic or optical variations present during manufacturing, preventing counterfeiters from replication~\cite{mcgrath2019puf}. \emph{Challenges} depend on the type of \ac{puf} and can be customized based on user-controlled parameters. By collecting a set of \acfp{crp} from a \ac{puf}, a product can be identified.

\acp{puf} rely on three key properties: reproducibility, unpredictability, and uniqueness~\cite{prada2020puf}. The former ensures that a \ac{puf} consistently generates the same response for the same challenge. Unpredictability ensures that different challenges yield uncorrelated responses. Lastly, uniqueness establishes that no two \acp{puf} generate the same response for the same challenge. Additionally, \acp{puf} are tamper-resistant, meaning they cannot be physically removed from a product without being destroyed.

To facilitate the data management for \ac{puf}-based identification, an
important goal is to reduce the required information to a constant
value, termed \emph{fingerprint}. As shown in the following, the fingerprint format strongly depends on the
type of \ac{puf}.

\subsubsection{Optical \ac{puf}}

Optical \acp{puf}'s leverage physical features like random placement of fluorescent nanowires~\cite{kim2014anti} or paper fiber structures~\cite{mcgrath2019puf}. These \acp{puf} often have limited numbers of \acp{crp}, sometimes just one~\cite{kim2014anti}. E.g., the nanowire \ac{puf} compares a scanned pattern to a previously captured image such that a corresponding fingerprint only has to hold a single image.

\subsubsection{Electronic \acp{puf}} Electronic \ac{puf} use the random physical properties of electronic circuits. This includes the behavior of ring oscillators~\cite{bossuet2013puf}, memory modules (cf. SRAM \ac{puf}~\cite{intrinsicid2020sram}), and the emission of radio signals~\cite{dejean2007rf}. It is common that the hardware that houses the \ac{puf} itself also records its responses~\cite{mcgrath2019puf}. When that is the case, additional interfaces are required to forward \ac{puf} responses to external verifiers. However, counterfeits can exploit this indirect \ac{puf} access by impersonating such interfaces and sending known \ac{puf} responses of original products. It follows that electronic \acp{puf} are restricted in their capability to publicly reveal responses. This problem is solved by using every \ac{crp} only once, requiring a trusted service that holds a set of enrolled \ac{crp} in a database~\cite{mcgrath2019puf}.

Methods eliminating the storage requirements of \ac{crp}
tables are based on \acf{zk} schemes where the same \ac{puf} response is reused
but never sent over any interface. Instead, such schemes only send a proof
of knowledge for a secret \ac{puf} response based on a previously published
\emph{commitment}~\cite{prada2020puf, iovino2019non, felicetti2023deep}.
Finally, the required storage space can also be reduced through \ac{puf}-based
hash chains, as proposed by Petzi~\cite{petzi2024puf}. Both, \ac{zk} commitments
and public hash chain values can be directly used as \ac{puf} fingerprint.

\section{Design}
\label{sec:design}

In this section we present \spc, a novel approach for prevention of counterfeiting that offers supply chain tracing and product lifecycle logging. Our approach addresses the limitations of related works by minimizing blockchain utilization, considering privacy requirements of involved parties, ensuring detailed traceability, and offering compatibility with various physical identification methods.

\begin{figure}[t]
\centering
    \includegraphics[width=\linewidth]{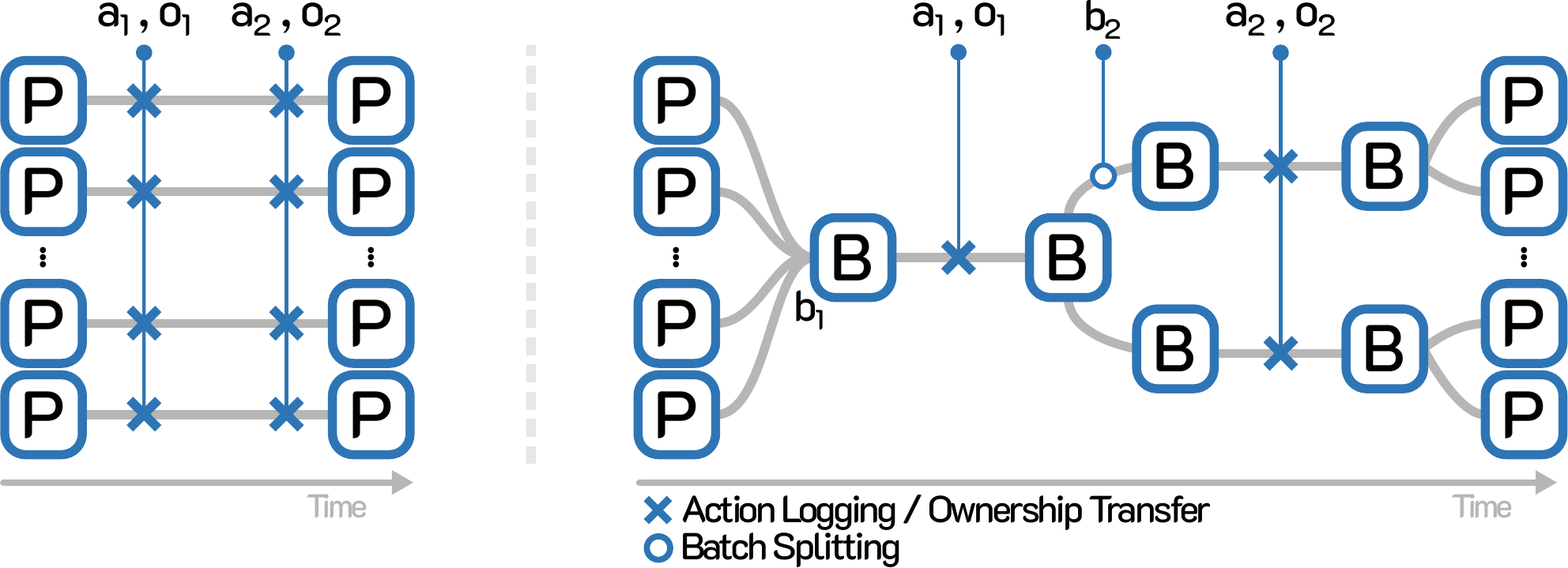}
    \caption{Product-wise (left) and batch-based (right) management of the
    lifecycle of a batch (B) of products (P).}
    \label{fig:eval:perf-2}
\end{figure}

\subsection{Requirements}
\label{sec:design:req}
In the following, we present the requirements for \spc. They are motivated by limitations of existing works, which we discuss in detail in Section~\ref{sec:relwork}. In particular, we find that existing systems neglect important privacy features and, in many cases, publish redundant information in the blockchain, limiting the scalability. Furthermore, the capability to log details of lifecycle events is not supported in the majority of existing systems.

\begin{renum}
\item Traceability: It must be possible to individually trace any products through the supply chain and log lifecycle events. Furthermore, tracing and event logging must be supported both within the supply chain and until a product's end of life.

\item Efficiency: The system should reduce the blockchain workload to overcome scalability limits of blockchains. For instance, the amount of data stored on blockchain should be minimized, as well as blockchain communication (i.e., transaction volume). 

\item  Privacy: Access to lifecycle data must not be public, but only available
to parties with justified interest. This implies that entities that publish
lifecycle events (e.g., supply chain actors or product owners) should be able to
(selectively) define, if their published event is to be accessible by other
entities up or down the supply chain.
Furthermore, relations
between users and products must not be publicly observable. E.g., it should not
be possible to establish the number of products that a producer publishes and it
should not be trivial to establish a link between the product and the identity
of its owner. 

\item Interoperability: The system must support various product identification methods to be able to accommodate the needs of various industries. E.g., the system must be able to trace both, electronic and non-electronic goods, and enable use of \ac{puf}-based identification methods, as well as more cost-effective (but less cloning-resilient) alternatives.
\end{renum}

\subsection{Key Design Features}

To fulfill requirements described in Section~\ref{sec:design:req}, \spc~relies on a number of design features, which we discuss in the following.

To achieve traceability (R1), products in \spc~are represented as virtual \emph{assets} that are published in a blockchain. In addition to holding a list of (past) owners, the blockchain is responsible for linking assets with subsequent lifecycle events. Such events, termed \emph{actions}, can be written by both, supply chain actors and end customers, to document, e.g., repairs, service, or returns. 

For reasons of efficiency (R2), multiple products can be represented by a \emph{batch} such that individual products do not have to
be published in the blockchain. This reduces the logging of similar lifecycle events that affect a set of products to a minimum (cf. the logging of similar actions $a_1$ and ownership transfers $o_1$ in Figure~\ref{fig:eval:perf-2}). Whenever needed, batches can be split into \emph{sub-batches} (cf. $b_2$ in Figure~\ref{fig:eval:perf-2}) to reflect real-world events such as unboxing of containers. Furthermore, to ensure (R1) fulfillment,  products that are contained in batches can be published on their own at any time (i.e., on demand) such that their lifecycle becomes independent of their parent batch. In cases where lifecycle
tracing ends at, e.g., a dealer, the publication of asset entries for individual products can be omitted, minimizing the workload for the utilized blockchain. This helps to reduce blockchain communication and on-chain storage. 

Further efficiency improvement is achieved by outsourcing data storage to
external and personally selectable storage systems that can be, e.g., hosted by
producers (e.g., cloud-based database), instantiated using distributed third
parties (e.g., \acf{ipfs} file systems~\cite{benet2014ipfs}), or be placed on \ac{nfc} tags attached
to a product. The blockchain storage is only utilized to store reference values
that help to verify integrity and authenticity of data records. 

To fulfill privacy requirements (R3), \spc~enforces access control to data held in storage systems, e.g., based on roles or
knowledge of secrets. This allows product owners to dynamically restrict the insight of other users into a product's lifecycle. Furthermore, users are allowed to create an abundance of blockchain addresses for anonymous interaction with different assets. E.g., producers can use this approach to hide the number of published batches.

Finally, to achieve interoperability (R4), \spc~defines a data format compatible with various identification methods. \spc~clients derive the required scanning hardware from the format and instruct users to perform the correct physical verification steps.

\begin{figure}[t]
\centering
    \includegraphics[width=0.9\linewidth]{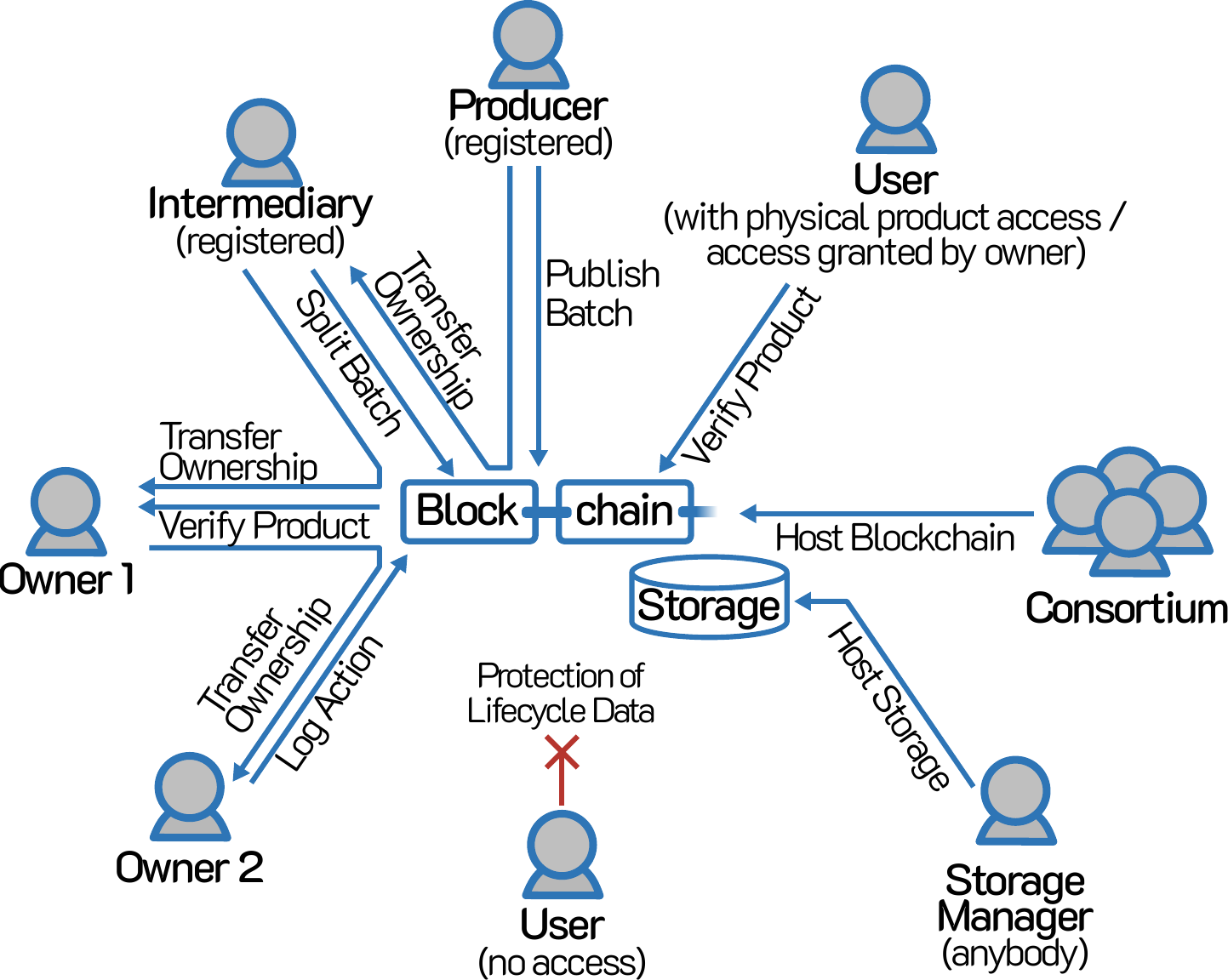}
    \caption{\spc~System Model}
    \label{fig:approach:overview}
\end{figure}
\subsection{System Model}
\label{sec:design:architecture}

The \spc~system model (cf. Figure~\ref{fig:approach:overview}) includes the entities: \emph{Consortium}, \emph{Producer}, \emph{Intermediary}, \emph{Owner} \emph{User}, and \emph{Storage Manager}. 

\begin{itemize}
    \item \textbf{Consortium}: \spc~leverages a permissioned blockchain, which is hosted by a \emph{consortium} -- a distributed trusted entity. Besides blockchain hosting, the consortium is also responsible for certain administration tasks, such as admitting new or revoking misbehaving actors. 
    \item \textbf{User}: Any entity such as end customer or supply chain actor
    that interacts with \spc~is considered to be a user. Every user can create
    one or more anonymous \emph{user addresses} that correspond to the public
    part of an asymmetric key-pair. Users can hold a combination of additional
    roles that are defined in the following.
    \item \textbf{Owner}: Users can be
    the (temporary) owner of an asset (batch or product). Owners are capable of
    transferring the ownership to other users, e.g., through a sale. Also,
    owners are capable of logging certain information about assets they own,
    e.g., log service events, or report issues with quality.
    \item \textbf{Producer}: Producers
    are product manufacturers that are registered in \spc~by enrolling with the consortium. Once registered, producers are
    associated with a \textit{user entry} on the blockchain and have the
    permission to publish new products and batches. The user entry holds
    their name, their manufacturer role, and the public part of a cryptographic key-pair for
    signature verification.
    \item \textbf{Intermediary}: Supply chain actors that do not create new
    products themselves are termed \emph{intermediary}. Similar to producers, intermediaries must be registered with the consortium and, once registered, are associated with a user entry. Intermediaries have the permission to split batches into sub-batches or publish their individual products.    
    
    \item \textbf{Storage Manager} With \spc, asset data is outsourced to various \emph{storage systems}. Any actor that hosts an external storage system
    and, thereby, has direct access to (a subset of) \spc's data is termed \emph{storage manager}.
\end{itemize}

\noindent
\textbf{Attacker model.} The \spc~system model acknowledges that attackers and
counterfeiters could have an interest in accessing confidential product
information and influencing the system or its users adversely. For instance, an
attacker could attempt to inject counterfeits into the blockchain records. In
addition to counterfeits, attackers could attempt to inject simulations, i.e.,
products that only look similar to the original but might, e.g., have a slightly
altered brand name~\cite{spink2013defining}. Furthermore, attackers could try to
claim ownership of products, e.g., in the case of theft. Finally, attackers could attempt to learn sensitive information, e.g., reveal how many products are produced by producers, or establish links in between products and identities of product owners.

To achieve one or more of these objectives, an attacker could corrupt any of the roles defined above, except for the \emph{consortium}, which is a (distributed) trusted party.
In case of dishonest consortium members, standard \ac{bft} consensus-based security assumptions ensure that fraudulent actions violating smart contracts are detected. Although this does not cover passive attacks such as eavesdropping, we perform a comprehensive security evaluation that assesses the (limited) impact of such attack.

\subsection{Product- and Batch-based Tracing}
\begin{figure}[t]
\centering
    \includegraphics[width=0.9\linewidth]{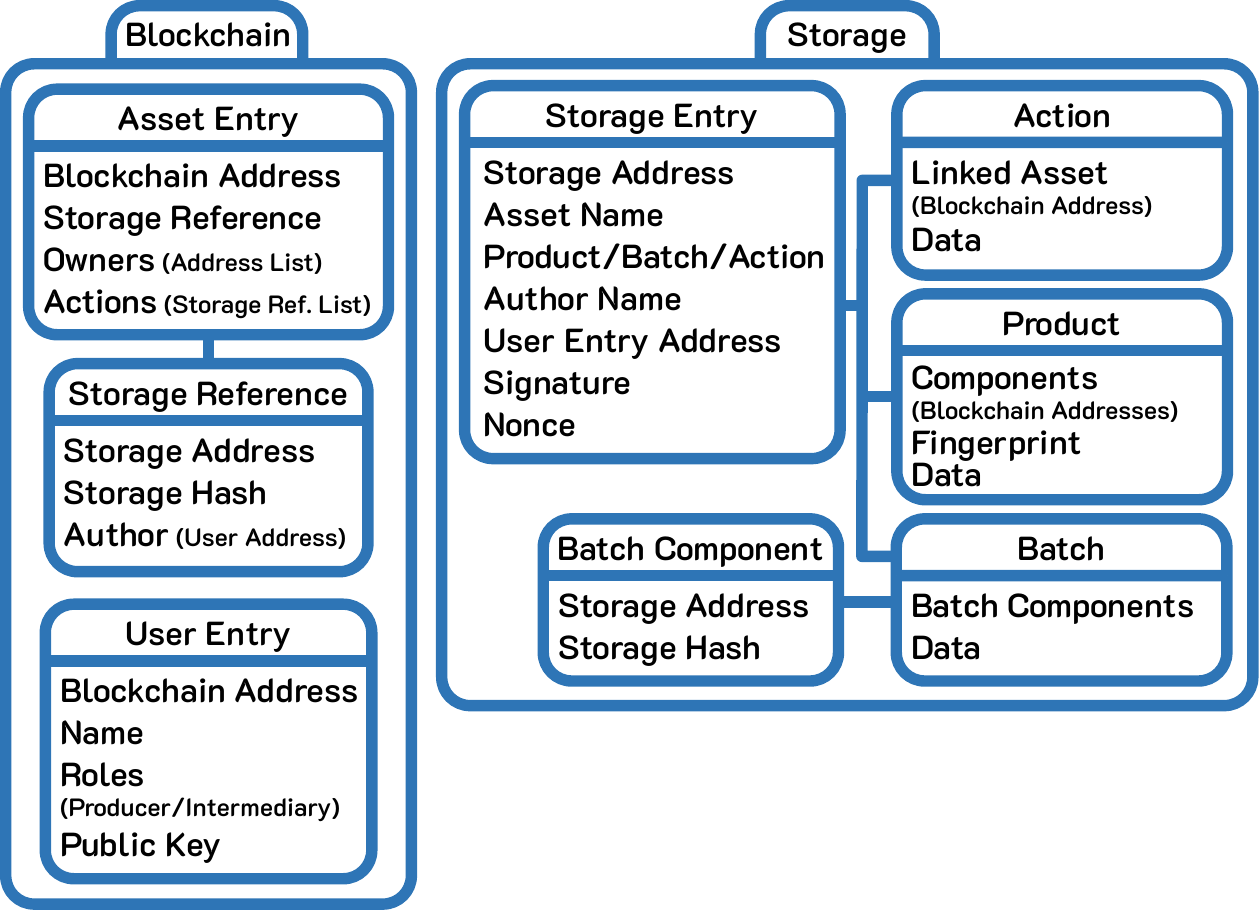}
    \caption{Asset, user, and storage entries in \spc.}
    \label{fig:design:uml}
\end{figure}
\spc~defines three asset types: \emph{Products}, \emph{Batches}, and
\emph{Actions}. For products and batches, an \emph{asset entry} can be published in the blockchain (cf. Figure~\ref{fig:design:uml}). Every asset entry is identified by a unique blockchain address. To acknowledge
lifecycle events that occur after asset publication, subsequent data can be
added to asset entries in the form of actions. For every asset type, the asset-specific information is outsourced to off-chain \emph{storage
entries} whose address and hash values are held in the blockchain.

Storage entries contain general information such as the asset name and author name (e.g., a brand name), as well as a customizable data field whose content can be freely selected. Every entry can be signed by producers and intermediaries to assert authorship. For products, the entry holds a
fingerprint and, optionally, the blockchain addresses of its components. For
batches, the entry holds a collection of \emph{batch components}. Every batch
component references the storage address and hash of either another batch or a
product. During publication of a batch, its components do not
yet have own asset entries in the blockchain. Finally, actions hold the blockchain address of the asset that they are associated with.
To prevent the prediction of an entry's contents based on its published hash value, every storage entry additionally holds a random nonce of 32 bytes.

\subsection{Product Identification}
\spc~is compatible with multiple types of product identifiers (termed \emph{fingerprint}), including barcodes, \ac{nfc} tags, and \acp{puf}. The identifier type is chosen by producers, e.g., based on the desired security level and in regard to the identifier's manufacturing costs.
Furthermore, producers account for the scanning hardware that their customers possess. E.g., if the product is sold to professionals, advanced \acp{puf} that require specialized tools such as industrial cameras could be utilized. In contrast, products that are sold to private customers should be equipped with established interfaces such as smartphone cameras and \ac{nfc} modules.
In all cases, the fingerprint field in the product's storage entry is designed to store information needed for performing the physical identification process.

\subsection{User-centric Privacy Controls}
\label{sec:design:privacy}
In case an asset's lifecycle consists of multiple logged actions,
product components, or parent batches, a holistic view of the lifecycle
requires access to the information of all related assets and their storage
entries. However, such comprehensive insight is potentially not mandatory. E.g.,
end customers could have no justifiable reason to access the information of
certain actions logged within the supply chain. To restrict access to such information, storage entries can enforce access control, as further described in Section~\ref{sec:design:storage}.

With this dynamic restriction of views on a product's lifecycle, producers and intermediaries can agree to use transparent lifecycle tracing within the supply chain but follow a black-box approach towards end customers. Similarly, the product's maintenance could be transparent to its actors but hidden from supply chain actors.

\begin{figure}[t]
\centering
    \includegraphics[width=\linewidth]{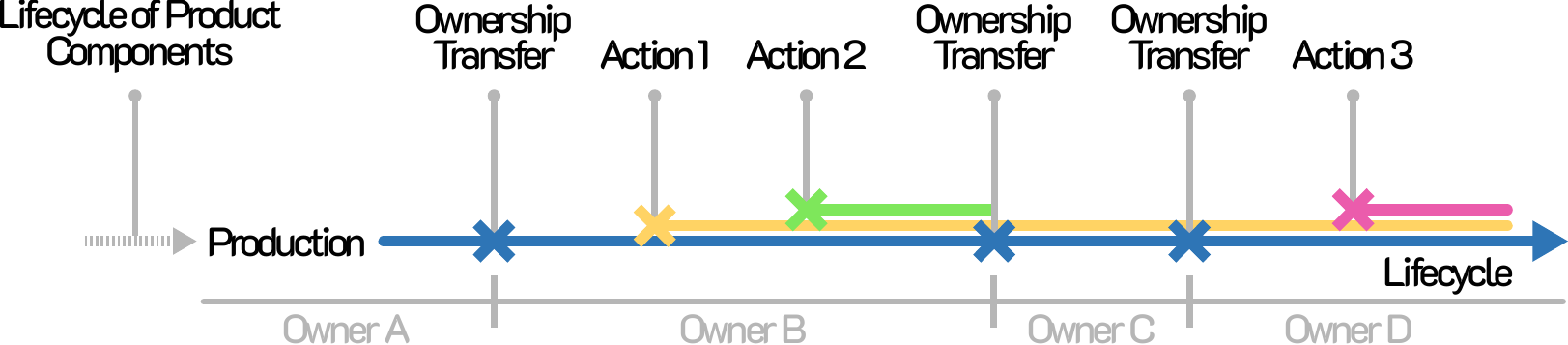}
    \caption{Exemplary product lifecycle (blue timeline) with restricted view
    on past and future events for different actors.}
    \label{fig:eval:lifecycleconcept}
\end{figure}

We illustrate this principle in Figure~\ref{fig:eval:lifecycleconcept}: Owner B creates two actions and gives owner C access to both, e.g., by sharing keys that are required for access (cf. Section~\ref{sec:design:storage}).
However, owner C decides to give owner D access
to only one action. In return, the subsequent action logged by owner D is not shared
with previous owners.

\begin{table}[b]
\centering
\begin{tabular}{r|c|c|c|c|c|}
\cline{2-6} &  \footnotesize Producer & \footnotesize Intermed. & \footnotesize Owner & \footnotesize User & \footnotesize Consortium \\
\hline
\multicolumn{1}{|r|}{} \footnotesize New asset entry & \checkmark & & & & \\
\multicolumn{1}{|r|}{} \footnotesize Publish component & & \checkmark & \checkmark & & \\
\multicolumn{1}{|r|}{} \footnotesize New sub-batch & & \checkmark & \checkmark & & \\
\multicolumn{1}{|r|}{} \footnotesize Log action & & & \checkmark & & \\
\multicolumn{1}{|r|}{} \footnotesize Transfer ownership & & & \checkmark & & \\
\multicolumn{1}{|r|}{} \footnotesize Query entries & & & & \checkmark & \\
\multicolumn{1}{|r|}{} \footnotesize Manage user entries & & & & & \checkmark \\
\hline
\end{tabular}
\caption{\spc~smart contract functions; Function calls are restricted to
    users that possess all marked roles}
\label{tab:design:smartcontract}
\end{table}

\subsection{Blockchain}
\label{sec:design:blockchain}

\spc~uses a permissioned blockchain for the smart contract-based management of
asset and user entries. 
In order to record asset lifecycles, users create and manage asset entries while the consortium is responsible for admitting new users. 
Therefore, the utilized smart contract ensures that user entries can only be modified by
the consortium while access to asset-related smart contract functions is
restricted based on user roles (cf. Table~\ref{tab:design:smartcontract}). E.g., only producers can create new asset entries for products and batches while only the consortium can create and modify user entries. Intermediaries can publish batch components and split batches into sub-batches, given that they own the associated batch. 
Finally, asset owners can perform basic lifecycle recording by logging new actions and passing the asset's ownership on to others. Users that do not own assets can only query the asset and user entries held by the blockchain.

It must be noted that access control for the roles of producers
and intermediaries is not performed for protecting the authenticity of asset
entries that these users can create: Independent of this access control, the
authenticity and authorship of assets is always verified based on the signature
that is held in their storage entry. Instead, this approach prevents that the
blockchain is overwhelmed with bogus asset entries that are submitted by unknown
actors.

To protect their privacy, the authentication of producers and intermediaries is
performed with ring signatures~\cite{rivest2001leak}. Those anonymize the authorship of signatures such that it can only be proven that they stem from a predefined
group of users (i.e., a collection of public keys). Still, they allow all
participating users to use individual private keys for signature
generation~\cite{rivest2001leak}. The concept has found application in
blockchains for increased privacy~\cite{li2020blockchain} and can be used within
smart contracts for role verification (i.e., membership of a group) of registered
\spc~users without revealing identities. The private keys
that are associated with \spc's user entries are used for creating public ring signature keys.

\subsection{Storage}
\label{sec:design:storage}

With \spc, asset data is held in decentralized storage systems. Any type of storage system (e.g., cloud storage or \acs{nfc}-tags) is supported as long as the
utilized \spc~client covers the corresponding methods for read/write access. Likewise, all users such as producers and intermediaries can host supported storage systems. All storage entries are identified by a unique storage address that is referenced, e.g., in asset entries in the blockchain. To identify the utilized storage system, the prefix of this address holds a descriptor that is unique for each supported storage system.

Storage systems must be accessible for all users that have a
legitimate interest in reading its storage entries. However, access control can
be applied to preserve privacy and share entries only with an intended set of
users. For this purpose, utilization of \textit{key-based} and \textit{ownership-based} access control for cloud-based storage systems are proposed. With the former, access is only granted with a pre-defined access key. In the latter case, the scheme verifies that the requesting user is the owner of the
associated asset. We present the details of each scheme in Appendix~\ref{app:access-control}.

Unregistered owners (e.g., end customers) can log actions without hosting storage, as the consortium is responsible for hosting a storage system with public write-access. To prevent misuse, ownership-based access control for write requests is applied. I.e., customers prove ownership of an asset entry before creating entries for associated actions.

\begin{figure}[t]
\centering
    \includegraphics[width=\linewidth]{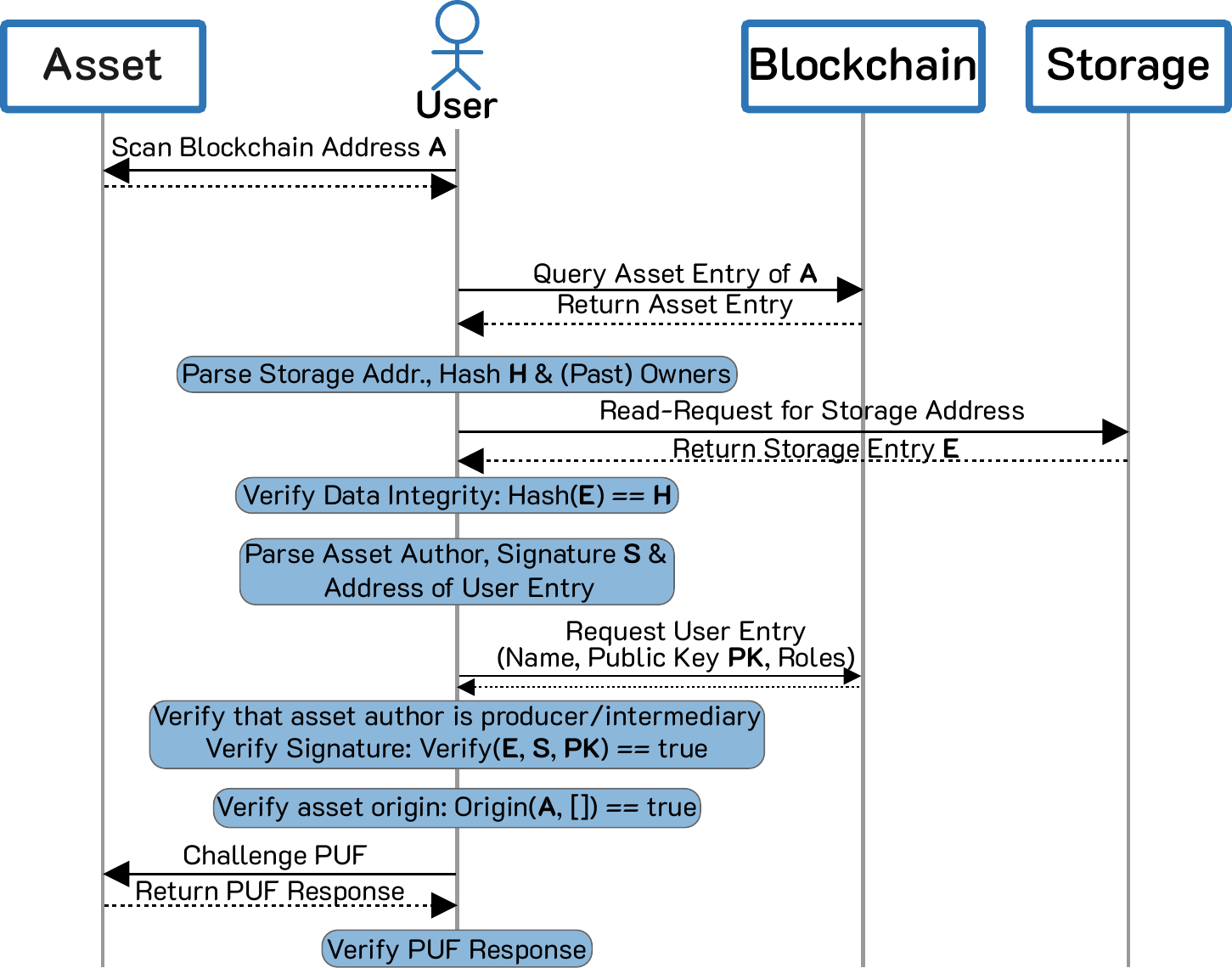}
    \caption{Asset verification.\\}
    \vspace{-0.5cm}
    \label{fig:design:asset-enrollment}
\end{figure}

\subsection{Use Cases}
\label{sec:design:use-cases}
This section elaborates on use cases for~\spc~and their connection to the previously introduced participants. As illustrated in Figure~\ref{fig:approach:overview}, all participants interact with~\spc~in specific ways.

\vspace{0.1cm}
\noindent \textbf{Preliminaries and user registration}
The consortium plays a crucial role in establishing the~\spc~infrastructure.
This includes hosting the blockchain, a core component of~\spc. Additionally, the consortium manages user registration, ensuring authorized access. Furthermore, the system relies on storage
systems that are provided by a designated role termed storage manager.

\vspace{0.1cm}
\noindent \textbf{Asset Publication}
Producers initiate product or batch registration within \spc~by creating
corresponding storage entries. Those hold critical data, including fingerprints and blockchain addresses of utilized components. For
batches, a components list is prepared with storage addresses and hashes of
sub-batches or products. Then, producers sign the storage entry and add it
to a chosen storage system. Finally, an asset entry is created with a blockchain
address that can be physically attached to the asset (e.g., QR code/NFC tag).

\vspace{0.1cm}
\noindent\textbf{Batch Splitting \& Batch Component Publication}
Intermediaries can generate new sub-batches that comprise a
selected subset of the parent batch's components. To split a batch, intermediaries create a new storage entry for each sub-batch, and invoke a smart contract function creating a new asset entry linked to the parent. Each sub-batch is assigned its own asset entry
that remains linked to the parent ensuring traceability. Further, intermediaries can publish existing batch
components within the blockchain, generating a distinct asset entry that is linked to
the parent batch's entry.

\begin{figure}[t]
\centering
    \includegraphics[width=0.9\linewidth]{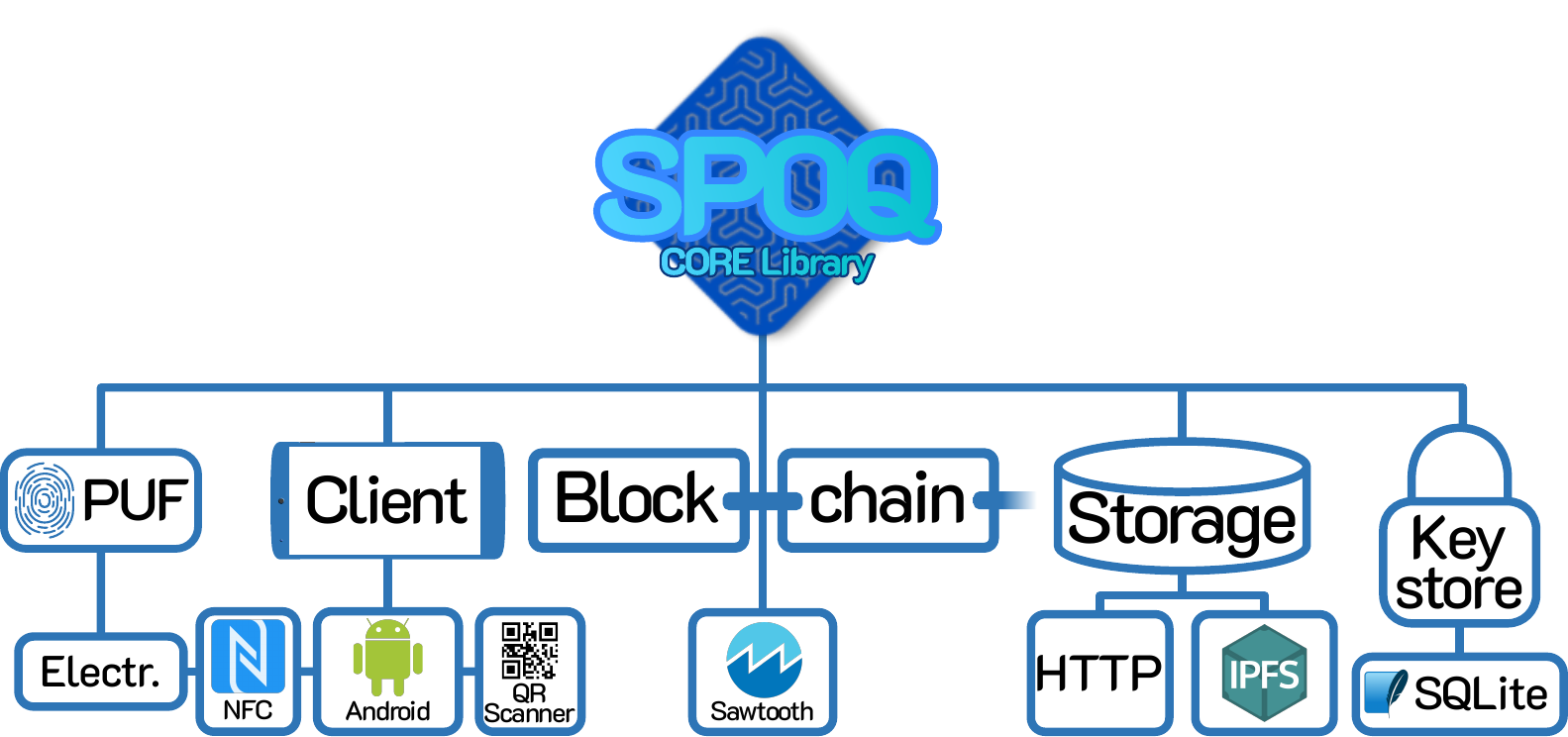}
    \caption{\spc~Software Components.}
    \vspace{-0.7cm}
    \label{fig:impl:arch-software}
\end{figure}

\vspace{0.1cm}
\noindent \textbf{Action Logging}
Asset owners can record lifecycle events by publishing asset-bound actions.
Initially, a storage entry is created that incorporates the action's data. If
the user is registered, the action can optionally be signed in order to assert authorship.
After publication of the storage entry,
the action's storage address and hash are added to the associated asset's list
of actions.

\vspace{0.1cm}
\noindent \textbf{Ownership Transfer}
To continue lifecycle tracing (i.e., log actions, split batches), ownership
of products and batches must be updated with each (legal) asset transfer. This process is
initiated by the recipient (i.e., buyer) that presents their user address,
e.g., in the form of a QR code. The current owner (i.e., seller) scans the
user address and appends it to the list of owners in the asset's
blockchain entry. The buyer can verify this transaction by querying the asset
entry. This process is (optionally) extended with the exchange of access keys that are
required for obtaining storage entries of related assets.

\vspace{0.1cm}
\noindent \textbf{Asset Verification} 
Users can retrieve information about products, batches, or actions that are stored
within~\spc. This process, outlined in Figure~\ref{fig:design:asset-enrollment},
begins by retrieving the blockchain address and required access keys, e.g., by scanning a
QR code that is attached to the asset. Subsequently, an automatic verification
process fetches relevant data from the blockchain and from storage entries that are linked with
the address.

To ensure data integrity and validity, several verification procedures
are conducted, as illustrated in Figure~\ref{fig:design:asset-enrollment}. The system
compares the storage entry's hash value with the one published on the blockchain
to detect any tampering since its creation. Additionally, the user that created
the entry is verified through a two-step process: The system checks the
the storage entry's signature using the author's public key that is obtained from the
blockchain. Furthermore,
the author's role within~\spc~is verified to confirm that they have the necessary
permissions for creating products or batches.
If the asset originates from a batch or is not even published in the blockchain on its own,
an additional step verifies the asset's
origin through its parent batch(es). An algorithm, described in detail in
Appendix~\ref{app:origin-verification}, ensures that the asset's trail can be
traced back to a root batch that is created by a producer. Finally, if the asset is
a product, users can perform a physical verification step using the provided
fingerprint. E.g., this can involve the use of an \ac{nfc} interface with a
compatible~\spc~client, assuming the user's device possesses the required
\ac{puf} scanner. Depending on the type of product, cheaper verification could be achieved by photographing a product's unique surface~\cite{mcgrath2019puf}.

If any verification step fails, the system assumes the asset's originality is compromised. Users are informed of issues such as
a failed signature verification, indicating unauthorized data modification.

\section{Implementation}
\label{sec:impl}

\spc~consists of multiple components, including a smart contract, storage systems,
and physical identification methods (cf. Figure~\ref{fig:impl:arch-software}). To facilitate
the interaction with all systems, a portable core library is implemented that is
used as the foundation for a mobile Android App but could also be
embedded into existing supply chain management systems. The size of all components is summarized in Table~\ref{table:impl:statistics} in regard to their lines of code.

\begin{table}[b]
    \centering
    \small
    \begin{minipage}{0.48\linewidth}
\hspace*{\fill}\begin{tabular}{|r|l|}
\hline
    Module          & \acs{loc}    \\
\hline\hline
    Core Library         & 15526                 \\
\hline
    Android Client       & 4070\\
\hline
    Sawtooth \acs{tp} & 784\\
\hline
\end{tabular}
    \end{minipage}
    \begin{minipage}{0.48\linewidth}
\begin{tabular}{|r|l|}
\hline
    Module          & \acs{loc}    \\
\hline\hline
    HTTP Server              & 566\\
\hline
    IPFS Gateway             & 193\\
\hline
    \ac{nfc} \acs{zk} \ac{puf} & 295 \\
\hline
\end{tabular}
    \end{minipage}
    \caption{This table shows the \acf{loc} for every module of \spc.}
    \label{table:impl:statistics}
\end{table}

\vspace{0.1cm}
\noindent\textbf{\spc~Library} The Go~\cite{golang} library defines the format of asset, user, and storage entries as well as \ac{puf} fingerprints. With over 15500 lines of code, the library implements all use cases of \spc~and defines standard
cryptographic functionality and serialization methods, ensuring compatibility
with external components that build on the library.
To facilitate the management of a user's owned and previously encountered assets, the library
includes an SQLite-based~\cite{gaffney2022sqlite} \emph{keystore} that locally stores the addresses and access keys of known assets.

\vspace{0.1cm}
\noindent\textbf{\spc~Front End} The implemented Android App acts as a front end for the \spc~library. It can be used by all \spc~users that, based on their role, can create new assets, transfer asset ownership, log actions, or verify an asset's originality. To scan blockchain addresses of assets and users, the App additionally offers a QR code scanner. Furthermore, it holds functionality for enrolling and verifying electronic \acp{puf} via \ac{nfc}.

\vspace{0.1cm}
\noindent\textbf{Product Demonstrator} To demonstrate \ac{puf}-based asset verification,
an electronic mockup \ac{puf} is developed that can be verified using \ac{zk}
proofs, following a scheme proposed in~\cite{felicetti2023deep}. This \ac{zk} \ac{puf} scheme is embedded into a Raspberry PI that uses a PN532 chip for
\ac{nfc}-based communication with the Android App. During the publication of new
products, the App queries the device for a \ac{zk} commitment that acts as
\ac{puf} fingerprint. To identify the verification type (i.e., electronic
\ac{zk} \ac{puf}, queried over \ac{nfc}), a unique header is prepended to the
fingerprint. For verification, the device is queried for a \ac{zk} proof.

\vspace{0.1cm}
\noindent\textbf{Blockchain} Our implementation leverages the permissioned Hyperledger Sawtooth blockchain for
the management of \spc's asset and user entries. With Hyperledger Sawtooth, smart contracts are termed \acf{tp} and manually installed at every node of the network.
The \ac{tp} for \spc~is written in Go and handles all function
calls related to asset and user entries.
Hyperledger Sawtooth allows \acp{tp} to interact with data entries that are
identified by a 35 bytes long address. This is used by \spc~to store and identify asset and
user entries of the format illustrated in Figure~\ref{fig:design:uml}.
For increased efficiency, the \ac{tp} is designed to accept multiple function calls within a single transaction. This allows, e.g., to submit multiple calls for batch splitting in one go, reducing the overhead induced by transaction headers.

\vspace{0.1cm}
\noindent\textbf{Storage Systems}
\label{sec:implementation:storage} The \spc~implementation includes two storage systems connected to
the Internet. For both systems, the \spc~library holds clients that enable read-
and write-access.

The first system, termed \emph{HTTP Server} is an access controlled database
that can be hosted freely by \spc~users such as producers and intermediaries
that wish to retain data ownership. The system's storage manager grants
write-access to a selected set of users. For reading storage entries, the key- and ownership-based access control schemes defined in
Appendix~\ref{app:access-control} are applied. To obtain storage entries, users must learn the system's domain name. To protect the storage manager's identity (as this could be, e.g., a producer that is also the asset author), the published storage address that is derived from the entry's hash value can be configured to not include the domain name. In this case, the domain name is shared with legitimate users via external channels, e.g., along with the sharing of the access key or during ownership transfers.

A second implementation is based on the decentral storage system termed
\ac{ipfs}~\cite{benet2014ipfs}. With \ac{ipfs}, nodes of a peer-to-peer network
combine their storage capacity. To serve as storage system, an Internet-facing
gateway is implemented that forwards read-/write-requests to an internal
\ac{ipfs} node (termed Kubo~\cite{ipfs-kubo}). Similar to the HTTP server,
\ac{ipfs} automatically derives storage addresses from the entry's hash value.
In the context of \spc, the consortium is responsible for hosting an \ac{ipfs}
network in order to allow, e.g., unregistered users without personal storage
systems to create new storage entries for logging actions. For increased
confidentiality, the client offers the option to encrypt storage entries before
publication. Similar to access keys, the encryption key can be shared with
other users via external channels.

\section{Evaluation of Efficiency}
\label{sec:efficiency}

In this section, we evaluate the efficiency of \spc. This is based on measuring its blockchain utilization and comparing it to product-wise tracing, as found in related work~\cite{shaffan2020blockchain,helo2020real,toyoda2017novel,caro2018blockchain}.

Both product-wise tracing and \spc~map
ownership transfers within the
supply chain to the blockchain. For comparison purposes, it is assumed that
existing systems are also capable of logging supply chain events as actions in
the blockchain. Although this feature is the fundamental basis of comprehensive
product lifecycle logging, it must be noted that only few existing systems
support action logging (cf. App. Table~\ref{tab:rel-work}).  Different to existing
approaches, \spc~enables the batching of products that are themselves traceable
without the need for immediate publication.
This principal difference is the subject of the evaluation performed in the
following.

\vspace{0.1cm}\noindent \textbf{Experiment}
Consider the production of a single batch ($b_1=1$) of $p$ products whose
observed lifecycle ends after their purchase through end customers, as illustrated in Figure~\ref{fig:eval:perf-2}. During
their supply chain traversal, a total of $a_1=10$ actions and
$o_1=10$ ownership transfers are registered for the original batch. Thereafter,
the batch is split into $b_2=10$ sub-batches for distribution among different
intermediaries. For each sub-batch, $a_2=10$ actions and $o_2=10$ ownership transfers
are registered. This scenario is illustrated in Figure~\ref{fig:eval:perf-2},
both for existing approaches that trace products individually (left) and for the
batch-based approach of \spc~(right). In the former case, the batch-wise ownership
transfers and logged actions must be repeatedly logged for every published product.

It follows that the number of transactions for product-wise tracing is computed
as
\begin{equation}
    \mathrm{tx}_\mathrm{EX} = p\cdot (1 + a_1+o_1+a_2+o_2)
\end{equation}
where the constant of $1$ represents the initial entry registration of each
product.
For \spc, the number of transactions is
\begin{equation}
    \mathrm{tx}_\mathrm{\spshort} = b_1\cdot (1 + a_1+o_1) +
    \left\{\begin{array}{ll}b_2 \cdot (1 + a_2+o_2) & \text{if } b_2\neq 0
        \text{,}\\
    b_1\cdot(a_2+o_2)&\text{else.}\end{array}\right.
\end{equation}
If sub-batching is not applied ($b_2 = 0$), the equation ensures that the
events $a_2$ and $o_2$ are counted in regard to the original batch.

In addition, the corresponding blockchain storage consumption at the lifecycle end is defined as
$\text{stor}_\mathrm{EX}$ and $\text{stor}_\mathrm{\spshort}$, respectively. 
This measurement is based on the observation that logged storage addresses and hashes have a length of $32$ bytes while user addresses consist of $33$ bytes. By using CBOR~\cite{bormann2020cbor} for serialization, an initial asset entry size of $\approx 182$ bytes~(no logged actions and only one owner) is obtained.

\begin{figure}[t]
\centering
\includegraphics[width=\linewidth]{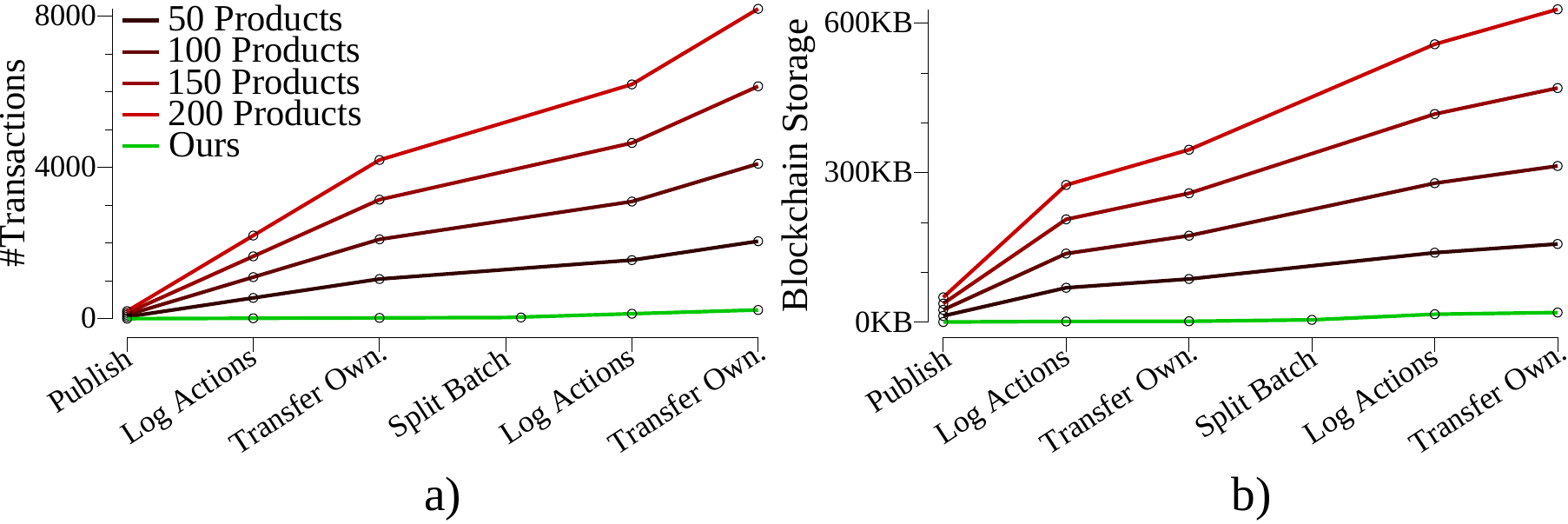}
\caption{
    Resource consumption for accumulating transactions (a) and blockchain storage (b) for product lifecycles in different
    publication modes. The lifecycles correspond to temporally consecutive events on the
    $x$-axis. The product-wise approach is evaluated for varying
    numbers of products.}
\label{fig:eval:compprodbatch}
\end{figure}

The lifecycle of a varying number of products in both existing approaches
(product-wise publication) and \spc~(batch-based publication) is modelled in
Figure~\ref{fig:eval:compprodbatch}. The rightmost values in
Figure~\ref{fig:eval:compprodbatch}a) correspond to the values
$\mathrm{tx}_\mathrm{EX}$ for $p\in\{50,100,150,200\}$ and
$\mathrm{tx}_\mathrm{\spshort}$, respectively. The associated storage
consumption is illustrated in Figure~\ref{fig:eval:compprodbatch}b).

\vspace{0.1cm} \noindent \textbf{Number of Required Transactions}
From Figure~\ref{fig:eval:compprodbatch}a), it is visible that redundant
action logging and ownership transferring, as performed with product-wise
tracing, has an unfavourable impact on
the number of required transactions. This can be compared to the number
of transactions that are required in \spc.

E.g., at $b_2=10$ sub-batches, it holds that \spc~requires $\approx 9$-times less transactions. Further results are listed in Table~\ref{tab:eval:theta}. By reducing the number of
sub-batches $b_2$ or eliminating batch splitting entirely ($b_2=0$), \spc~becomes $p$-times more efficient.

\begin{table}[b]
\centering
\begin{tabular}{|r|c|c|c|c|}
\hline
    & \multicolumn{2}{c|}{$\#$Transactions } &
    \multicolumn{2}{c|}{Storage Consumption } \\
 \hline
   $p$ & $b_2 = 10$ & $b_2 = 0$ & $b_2 = 10$ & $b_2 = 0$ \\
\hline
\hline
    $50$ & $9$  & $50$ &  $8$ & $50$ \\
    \hline
    $100$ & $18$ & $100$ & $16$ & $100$ \\
    \hline
    $150$ & $27$ & $150$ & $24$ & $150$ \\
    \hline
    $200$ & $35$ & $200$ & $34$ & $200$ \\
\hline
\end{tabular}
\caption{Efficiency factor of \spc~over product-wise tracing, measured in the proportion of
    storage space and the number of
    transactions}
\label{tab:eval:theta}
\end{table}

\vspace{0.1cm} \noindent \textbf{Storage Requirements} In regard to storage requirements, the most significant impact  is caused by the logging
of actions. Transferring ownership only adds a new
owner address to an existing entry and has limited impact on storage
requirements. Overall, the lifecycles of $200$ products require $628$KB while the batch-based approach results in $19$KB of storage consumption (cf. Figure~\ref{fig:eval:compprodbatch}b)), indicating $33$-times smaller impact on blockchain storage. The comparison with all product volumes is summarized in Table~\ref{tab:eval:theta}.
Again, this efficiency gain is decisively increased to $p$ by reducing the number of
sub-batches (cf. Table~\ref{tab:eval:theta}).

\begin{figure}[t]
\centering
    \includegraphics[width=\linewidth]{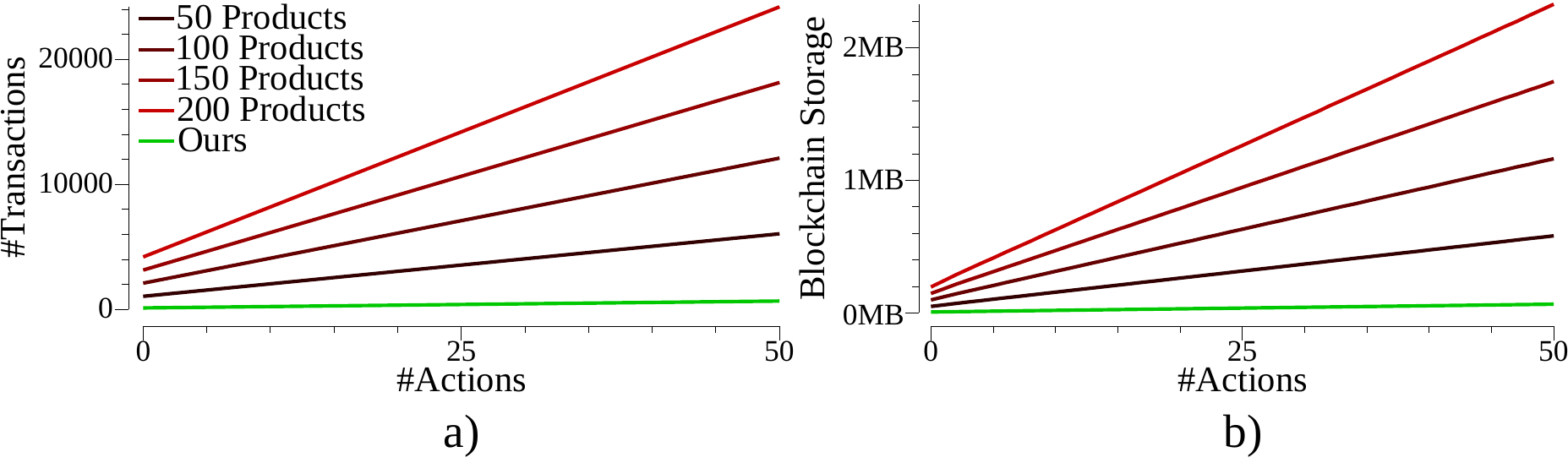}
    \caption{Resource consumption in number of transactions (a) and blockchain storage (b) for full asset lifecycles at different numbers of actions. The
    $x$-axis corresponds to $x=a_1=a_2$.
    }
\label{fig:eval:compprodbatch-numactions}
\end{figure}

\vspace{0.1cm} \noindent \textbf{Influence of Different Numbers of  Actions}
Based on the observation that action logging has the most significant increase of
storage space consumption, the experiment described above is
repeated for varying numbers of logged actions for each lifecycle.
The corresponding transactions and storage space requirements at the end of the lifecycle are illustrated in
Figure~\ref{fig:eval:compprodbatch-numactions}.

At $a_1=a_2=50$, the product-wise approach with $p=200$ has created $24200$
transactions and requires $2.3$MB of storage. In contrast, the batch-based
approach with $b_2=10$ sub-batches has created $671$ transactions, consuming 
$66$KB of storage. At $b_2=0$, i.e., when no batch splitting is performed, only
$121$ transactions are created that require $12$KB of storage.

\vspace{0.1cm}
\noindent \textbf{Summary} From both the number of transactions and the required storage space, it can be
observed that the efficiency is linearly proportional to the number of products
and sub-batches. If no sub-batching is applied, \spc~is more efficient with a
factor of $p$. I.e., the existing approach of product-wise tracing is linearly
less efficient than \spc.

\section{Security Analysis}
\label{sec:eval}
This section evaluates \spc's privacy and security guarantees. In particular, we informally argue how security and privacy requirements formulated in Section~\ref{sec:design:req} are fulfilled in regard to the attacker model described in Section~\ref{sec:design:architecture}.

\subsection{Privacy Review}
\label{sec:eval:privacy}
To assess privacy guarantees, we discuss the public and confidential information stored and maintained by \spc. We then elaborate on how privacy is achieved for privacy-sensitive data, such as owner identifiers and asset entries. 

\subsubsection{Private Asset Data}
In \spc, asset data is stored in a (user's personally selected and trusted) storage system
that protects entries based on the preferred method of access control. E.g., the HTTP server implemented in Section~\ref{sec:impl} protects access either based on secret keys or based on ownership verification. In contrast, the distributed \ac{ipfs} can protect access through encryption. It holds that asset data is only accessible to trusted storage managers and to users that the asset's author has shared access with.

\subsubsection{Publicly Accessible Data}
\label{sec:eval:privacy:public-data}
The only publicly accessible information is found in the asset and
user entries held in the blockchain. In the following, we discuss the information that is exposed in the public data entries for users and assets.

\vspace{0.1cm}
\noindent
\textbf{User Entries} 
User entries contain name and associated public key of producers and intermediaries along
with their roles within \spc. It is assumed that users provide their public business name such that this information does not reveal private identities.

\vspace{0.1cm}
\noindent
\textbf{Asset Entries} The data stored in publicly accessible asset entries is listed in
Figure~\ref{fig:design:uml}. Besides a specification of the asset type, the
entry also contains a list of user addresses of past owners. Those addresses, however, are not associated with a user's identity. E.g., a user can choose to use individual user addresses for different assets. Finally, the entry also
holds storage related information for the asset and its linked actions,
consisting of a storage address and a data hash. The data hash is protected with a random nonce, impeding the capability to brute force the entry's contents or compare it to other entries. For the two implemented storage systems, it holds that the storage address is also derived from the entry's hash and that no additional information is learned from the address except for the utilized type of storage system.

\subsubsection{Consortium} Consortium members have access to transactions and network traffic in the permissioned blockchain. With this data,
consortium members can, e.g., relate IP
addresses of users to asset entries.
However, users can apply established methods, e.g.,
virtual private networks for hiding their true address.

As shown in Section~\ref{sec:design:blockchain}, anonymous blockchain
interaction for producers and intermediaries is achieved by the use of ring
signatures. This approach disallows the consortium to link asset entries with
user identities. Concluding, it is possible to operate on \spc~anonymously without
leaking asset-user-relations, e.g., by using anonymous IP addresses
and ring signature-based authentication methods.

\subsubsection{Batches}
The storage entries of
batches list assets that a user may not be allowed to
access. I.e., if a user buys a product and obtains access to the batches
that this product was part of, the user also gains knowledge of the storage
entries of related assets
that are direct children of these batches. However, with access control applied on
these
storage entries, the user neither learns the type of asset (product or
sub-batch) nor their content. Without further knowledge of the batch's
structure, the user, therefore, cannot estimate the batch's true size. By
assuming that the corresponding producer utilizes multiple anonymous user addresses
for publishing batches,
it is also not possible to derive, e.g., the production volume.

\subsection{Security Analysis}
\label{sec:eval:attacks}
This section assesses the security of \spc~in regard to the attacker types
described in Section~\ref{sec:design:architecture}. Particularly, we discuss
attacks on the integrity of \spc~and the injection of counterfeits.

\subsubsection{Injection of Counterfeits} Unregistered counterfeiters do not have the capability to create
asset entries in the blockchain. Still, counterfeiters can try to bribe registered
users into asset registration by proxy. Producers can perform such registration
unhindered, whereas intermediaries can only register forged asset entries
through sub-component publication from a batch in their possession. In the
latter case, such action raises suspicion when potential customers verify the
origin trail of the counterfeit, given that the forged asset links to an
unrelated batch. Further, the asset verification
process automatically assesses the provided signature. Without having the
private signature keys for a valid user entry, counterfeiters cannot
produce valid signatures with the correct brand name for their forged asset data. Counterfeits are, therefore, immediately
detected as such whenever a client enrolls its associated asset.

\subsubsection{Simulation}
In contrast to counterfeits that are exact copies, simulations only \textit{look like} the original~\cite{spink2013defining}. E.g., the counterfeit could have a slightly altered brand name. In this scenario, counterfeiters can try to
become registered in \spc~with a slightly altered name, allowing them
to produce valid signatures while relying on the incapability of users to
detect the alteration. This issue can be ruled out through the consortium that
verifies a brand's authenticity during the registration process.

\subsubsection{Corruption}
\label{sec:eval:sec:corruption}

As previously described, registered users can abandon their trustworthiness
by being bribed by counterfeiters. A broader definition of corruption includes
illicit behavior such as publishing incorrect asset data, selling products
below advertised quality, or failing to follow pre-agreed protocols, or
forging the entries of bach-components during batch splitting.
While the latter can be automatically detected by \spc~clients, the product-specific misbehavior must be detected by, e.g., end customers, who can report such incidents to the
consortium. The
consortium, in turn, has the obligation to investigate and potentially withdraw
the permissions of the corrupted user. With their user entry being revoked,
the corrupt user loses all access to the \spc~infrastructure and all affected
victims can be informed. It follows that a single detection of corrupt behavior
is connected with high economic risk and can effectively prevent future
misconduct.

\subsubsection{Cover-Up}
\label{sec:eval:sec:cu}
Storage
managers can have a malicious intent to forge storage entries of
assets that are already registered in \spc's blockchain.
E.g., producers that host own storage systems can try to cover up
product-related incidents by changing the previously published data without
public notice. Due to the initial publication of the storage entry's hash value
on the associated blockchain entry, however, the probability of such cover-up
remaining undetected is bound to finding hash collisions.

\subsubsection{Ownership Fraud / Theft}
\label{sec:eval:sec:originf}
Individuals can claim properties of an asset's origin and ownership that do not
match the information logged via \spc. E.g., dishonest supply chain actors can
claim current ownership of an asset. However, verifying users can query the asset's
blockchain entry for the current owner and demand the actor to prove that the
linked blockchain address is in their possession.
In the case of theft, the legitimate owner
that the asset was stolen from can use \spc~to publicly announce the crime in
the form of an action that is logged in the asset entry.

\section{Related Work}
\label{sec:relwork}

To present related work, we categorize it into four key areas: Tracing,
efficiency, privacy, and identification. These categories are reflected in App. Table~\ref{tab:rel-work}, showing capabilities of related works
compared to \spc.

\vspace{0.1cm} \noindent \textbf{Tracing} The tracing group identifies three key
properties: Lifecycle events, individual goods, and batched tracing. Lifecycle
events support signifies the ability to track arbitrary events beyond ownership
transfers. The individual goods capability indicates that the system is designed
to trace single items. Finally, batched tracing allows efficient handling of
large numbers of products by grouping them, reducing processing overhead. Only
one related approach~\cite{abeyratne2016blockchain} fulfills all three
properties by proposing the creation of comprehensive product profiles that
support the logging of lifecycle events and association with further profiles.
However, the system assumes that all products are individually registered upon their creation. Furthermore, batches are designed to forward lifecycle events to their contained products, leading to high redundancy. The majority of works~\cite{shaffan2020blockchain, helo2020real, toyoda2017novel,
abeyratne2016blockchain, caro2018blockchain, baralla2019ensure,
marchese2021agri, aniello2021anti, negka2019employing, kennedy2017enhanced,
felicetti2023deep, westerkamp2018blockchain} focus on individual products. In contrast, the works~\cite{caro2018blockchain,baralla2019ensure,marchese2021agri,westerkamp2018blockchain} perform a batch-based tracing with the special characteristic that batches only specify a quantity of products of the same type.
With this mechanism, identities of product components are lost during
production. Further, all works~\cite{caro2018blockchain, baralla2019ensure, marchese2021agri, westerkamp2018blockchain} focus on specific industrial fields such as
agriculture and define corresponding smart contract functions (e.g., for
planting, growing, and harvesting), impeding immediate adoption by different
supply chains and limiting privacy in regard to data shared with blockchain
hosts. In contrast, \spc~fulfills all three properties allowing deployments in various fields. Further, related approaches lack support for
important properties in regard to efficiency, privacy, and identification, as we
will elaborate below.

\vspace{0.1cm} \noindent \textbf{Efficiency} The efficiency group focuses on two
key properties: Reducing blockchain transactions and enabling off-chain data
storage. Firstly, on demand publication eliminates redundant transactions,
specifically those related to registering products. This significantly improves
the overall efficiency of the approach. Secondly, off-chain data storage support
further enhances efficiency. By leveraging storage systems instead of the
blockchain for specific data, the system gains significant scalability. Related
approaches such as~\cite{shaffan2020blockchain, helo2020real, toyoda2017novel,
abeyratne2016blockchain, caro2018blockchain, baralla2019ensure,
marchese2021agri, aniello2021anti, negka2019employing, kennedy2017enhanced,
westerkamp2018blockchain} register all products in the blockchain, leading to
high demands for blockchain throughput and storage that is aggravated when
similar lifecycle events are logged multiple times for each product. The
approach from Felicetti\etal~\cite{felicetti2023deep} supports that product data
is outsourced to off-chain storage. However, the approach neither specifies
means for protecting data from illegitimate access nor support for the logging
of lifecycle events. Our approach fulfills all two efficiency properties.

\vspace{0.1cm} \noindent \textbf{Privacy} The privacy group encompasses three
aspects. First, data ownership extends the notion of off-chain storage in the
sense that users can keep data at their own servers. Second, the property of
protected data access ensures that only authorized users within the system can
access confidential lifecycle information.
Finally, hidden product-user
relationships indicate that the identities of product owners cannot be derived
from publicly accessible data. A majority of proposed
approaches~\cite{shaffan2020blockchain, helo2020real,
toyoda2017novel, caro2018blockchain, baralla2019ensure, marchese2021agri,
aniello2021anti, negka2019employing, kennedy2017enhanced,
westerkamp2018blockchain} store user and product data transparently. Due to the
absence of a permission model, records can be accessed at will.
In~\cite{abeyratne2016blockchain}, product profiles in the blockchain are
protected with access control. However, the data is still shared with all
blockchain hosts. \spc~is the only solution that fulfills all three privacy
properties by utilizing off-chain storage with access control and anonymous
blockchain addresses.

\vspace{0.1cm} \noindent \textbf{Identification} The identification group
addresses the use of physical technology for anti-counterfeiting, such as optical
and electronic \acp{puf}. The majority of supply chain tracing
systems~\cite{shaffan2020blockchain, helo2020real, toyoda2017novel,
abeyratne2016blockchain, caro2018blockchain, baralla2019ensure,
marchese2021agri, westerkamp2018blockchain} rely on insecure identification
methods, offering no protection against counterfeiting. The systems by
Aniello\etal\cite{aniello2021anti} and Negka\etal\cite{negka2019employing} are
tailored for \ac{ic}-based products, prohibiting support for analog products.
Similarly, the system by Kennedy\etal\cite{kennedy2017enhanced} relies on a
specific, fluorescence-based \ac{puf}, limiting interoperability. Greater interoperability can be expected for the \ac{nfc} tag-based
system by Felicetti\etal\cite{felicetti2023deep} that is extended with optical
surface detection. Although all described \ac{puf}-enabled systems offer
sophisticated counterfeit detection, they lack the interoperability with
different \ac{puf} types and the capability to comprehensively record a
product's lifecycle events. \spc~is the only solution that abstracts the
functionality of \acp{puf} and, thereby, offers producers to freely choose a
desired identification method.

\vspace{0.1cm} \noindent In summary, \spc~provides a comprehensive supply chain
tracing solution that surpasses existing approaches. It achieves this by
providing superior capabilities in traceability, efficiency, privacy, and
identification. Specifically, \spc~enables efficient tracing of individual goods
and batches. It also reduces blockchain load through non-redundant lifecycle
recording and leverages off-chain storage for scalability. In addition,
\spc~prioritizes privacy by supporting off-chain storage with access control and
anonymized product-user relationships. Finally, \spc~provides compatibility with
multiple product identification technologies.

\section{Conclusion}
\label{sec:conc}

This work introduces \spc, a novel supply chain tracing and counterfeit
protection platform that covers complete product lifecycles and, thereby, is an
ideal solution both for supply chain actors and end customers. Its focus lies
on high user privacy, system security, scalability, and compatibility with
diverse identification methods. \spc~reduces the redundant logging of similar
product lifecycles through the introduction of a sophisticated batching
mechanism. With this work, the design of \spc~is introduced, and its security,
privacy, and efficiency gains are successfully evaluated. Furthermore, the
provided implementation constitutes a highly adaptable, efficient, and
privacy-preserving platform that future applications can be built on.

\begin{acronym}
    \acro{b2b}[B2B]{Business-to-Business}
    \acro{b2c}[B2C]{Business-to-Consumer}
    \acro{bft}[BFT]{Byzantine Fault Tolerant}
    \acro{c2c}[C2C]{Consumer-to-Consumer}
    \acro{ca}[CA]{Certificate Authority}
    \acro{cdn}[CDN]{Content Delivery Network}
    \acro{cft}[CFT]{Crash Fault Tolerant}
    \acro{cid}[CID]{Content Identifier} 
    \acro{crp}[CRP]{Challenge-Response Pair}
    \acro{dapp}[DAPP]{Decentralized Application}
    \acro{ddos}[DDoS]{Distributed Denial of Service}
    \acro{dlt}[DLT]{Distributed Ledger Technology}
    \acro{dos}[DoS]{Denial of Service}
    \acro{dpp}[DPP]{Digital Product Passport}
    \acro{fr}[FR]{Functional Requirement}
    \acro{hce}[HCE]{Host-based card emulation}
    \acro{ic}[IC]{Integrated Circuit}
    \acro{iot}[IoT]{Internet of Things}
    \acro{ipfs}[IPFS]{InterPlanetary FileSystem}
    \acro{iqr}[IQR]{Interquartile range}
    \acro{loc}[LOC]{Lines of Code}
    \acro{ml}[ML]{Machine Learning}
    \acro{nfc}[NFC]{Near Field Communication}
    \acro{pbft}[PBFT]{Practical Byzantine Fault Tolerance}
    \acro{pki}[PKI]{Public Key Infrastructure}
    \acro{poet}[PoET]{Proof of Elapsed Time}
    \acro{poms}[POMS]{Product Ownership Management System}
    \acro{pow}[PoW]{Proof of Work}
    \acro{pps}[PPS]{Payloads Per Second}
    \acro{puf}[PUF]{Physical Unclonable Function}
    \acro{rfid}[RFID]{Radio-Frequency Identification}
    \acro{rq}[RQ]{Research Question}
    \acro{sbc}[SBC]{Single Board Computer}
    \acro{sdk}[SDK]{Software Development Kit}
    \acro{spoqen}[SPOQ]{Standardized secure Product verification for the protection of Originality and Quality}
    \acro{spoq}[SPOQ]{Standardisierte sichere Produktverifizierung zum Schutz von Originalität und Qualität}
    \acro{sr}[SR]{Security Requirement}
    \acro{tee}[TEE]{Trusted Execution Environment}
    \acro{tps}[TPS]{Transactions Per Second}
    \acro{tp}[TP]{Transaction Processor}
    \acro{wots}[W-OTS]{Winternitz One-Time Signature}
    \acro{wp}[WP]{Work Package}
    \acro{wto}[WTO]{World Trade Organization}
    \acro{zk}[ZK]{Zero Knowledge}
\end{acronym}


\begin{acks}
This research was funded by the WIPANO program of the German Federal Ministry for Economic Affairs and Climate Action, project SPOQ
(Standardized secure Product verification for the protection of Originality and Quality).
\end{acks}

\bibliographystyle{ACM-Reference-Format}
\bibliography{bib}

\appendix

 \begin{table*}[h]
 \begin{tabular}{@{}l|c|c|c|c|c|c|c|c|c|c|@{}}
 \cline{2-11}

                                & \multicolumn{3}{c|}{\textbf{Tracing}}                                                                                                                    & \multicolumn{2}{c|}{\textbf{Efficiency}}                                                                                                       & \multicolumn{3}{c|}{\textbf{Privacy}}                                                                                                                                                                    & \multicolumn{2}{c|}{\textbf{Identification}}                                              \\ \cline{2-11}\vspace{-0.15cm}
 & \rotatebox{90}{\parbox{1.9cm}{Lifecycle Events}} & \rotatebox{90}{\parbox{1.9cm}{Individual Goods}} & \rotatebox{90}{\parbox{1.9cm}{Batched\\Tracing}} & \rotatebox{90}{\parbox{1.9cm}{On Demand \\Publication}} & \rotatebox{90}{\parbox{1.9cm}{Off-Chain Data\\Storage}} & \rotatebox{90}{\parbox{1.9cm}{Data\\Ownership}} & \rotatebox{90}{\parbox{1.9cm}{Protected\\Data Access}} & \rotatebox{90}{\parbox{1.9cm}{\parbox{1.9cm}{Hidden Product-User-Relations}}} & \rotatebox{90}{\parbox{1.9cm}{Optical\\PUF}} & \rotatebox{90}{\parbox{1.9cm}{Electronic\\PUF}} \\ \midrule
 Shaffan\etal\cite{shaffan2020blockchain}             &                                                  & \checkmark                                       &                                                  &                                                            &                                                         &                                                 &                                                        &                                                                               &                                              &                                                 \\
 Helo\etal\cite{helo2020real}                         &                                                  & \checkmark                                       &                                                  &                                                            &                                                         &                                                 &                                                        &                                                                               &                                              &                                                 \\
 Toyoda\etal\cite{toyoda2017novel}                    &                                                  & \checkmark                                       &                                                  &                                                            &                                                         &                                                 &                                                        &                                                                               &                                              &                                                 \\
 Abeyratne \& Monfared~\cite{abeyratne2016blockchain} & \checkmark                                       & \checkmark                                       & (\checkmark)                                       &                                                            &                                                         &                                                 & (\checkmark)                                           &                                                                               &                                              &                                                 \\
 AgriBlockIoT~\cite{caro2018blockchain}               & \checkmark                                       & \checkmark                                       &                                                  &                                                            &                                                         &                                                 &                                                        &                                                                               &                                              &                                                 \\
 Baralla\etal\cite{baralla2019ensure}                 & \checkmark                                       &                                                  & \checkmark                                       &                                                            &                                                         &                                                 &                                                        &                                                                               &                                              &                                                 \\
 Marchese\etal\cite{marchese2021agri}                 &                                                  &                                                  & \checkmark                                       &                                                            &                                                         &                                                 &                                                        &                                                                               &                                              &                                                 \\
 Westerkamp\etal\cite{westerkamp2018blockchain}       &                                                  &                                                  & \checkmark                                                  &                                                            &                                                         &                                                 &                                                        &                                                                               &                                              &                                                 \\
 Anti-BlUFf~\cite{aniello2021anti}                    &                                                  & \checkmark                                       &                                                  &                                                            &                                                         &                                                 &                                                        &                                                                               &                                              & \checkmark                                      \\
 Negka\etal\cite{negka2019employing}                  &                                                  & \checkmark                                       &                                                  &                                                            &                                                         &                                                 &                                                        &                                                                               &                                              & \checkmark                                      \\
 Kennedy\etal\cite{kennedy2017enhanced}               &                                                  & \checkmark                                     &                                                  &                                                            &                                                         &                                                 &                                                        &                                                                               & \checkmark                                   &                                                 \\
 Felicetti\etal\cite{felicetti2023deep}               &                                                  & \checkmark                                       &                                                  &                                                            & \checkmark                                              &                                                 & (\checkmark)                                           &                                                                               & \checkmark                                   & \checkmark                                      \\
 \textbf{\spc}                                        &  \checkmark                                                & \checkmark                                      & \checkmark                                       & \checkmark                                                 & \checkmark                                              & \checkmark                                      & \checkmark                                            & \checkmark                                                                    & \checkmark                                   & \checkmark                                      \\ \hline
 \end{tabular}
 \caption{Properties of related supply chain tracing and counterfeit detection solutions}
 \label{tab:rel-work}
 \end{table*}

\section{Storage System Access Control}
\label{app:access-control}

This section proposes two access control schemes for asset data in storage systems.
The presented schemes are relevant primarily for systems that are publicly
accessible, e.g., from the Internet.

\subsection{Access Keys}
This scheme allows any user with knowledge of a storage entry's \textsf{access
key} to obtain the entry. An access key is defined to be a random value of 32
bytes that is set by the asset's author.
To facilitate asset enrollment by foreign users, the key can be shared along with an asset's blockchain address, e.g., through a QR code that is attached to a product or batch.

\begin{figure}[t]
\centering
    \includegraphics[width=\linewidth]{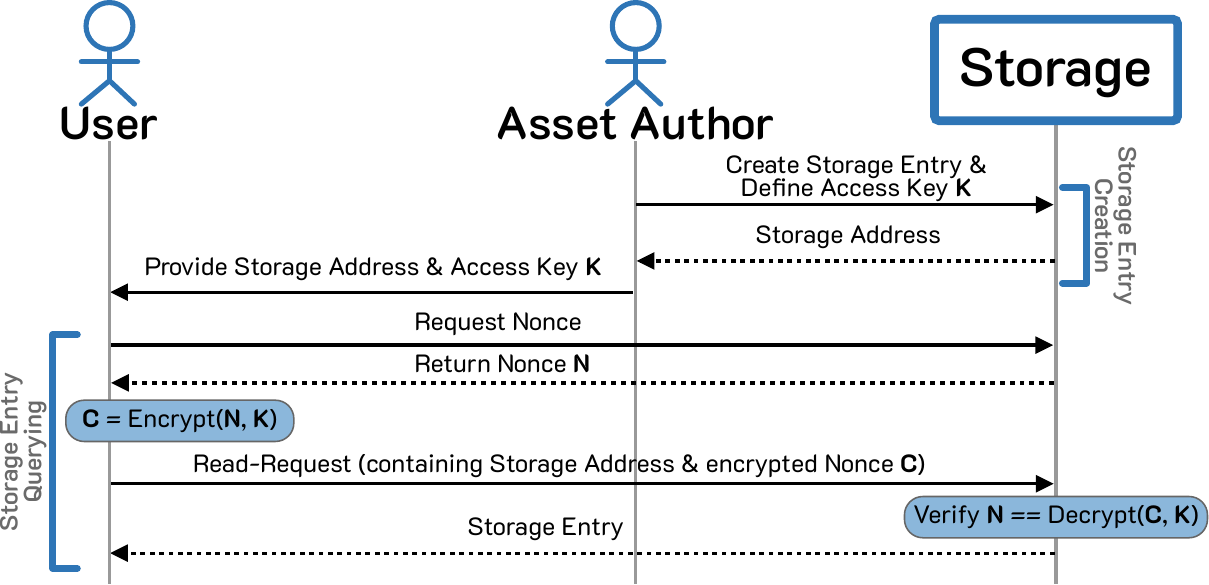}
\caption{Access key-based access control for storage entries.}
\label{fig:design:tokenauth}
\end{figure}

During storage entry registration, the author submits both the entry and
an access key $k$ to the storage system (cf.
Figure~\ref{fig:design:tokenauth}).
After receiving the storage address and access key, the user requests a nonce
$n$ from the storage system and encrypts it using symmetric algorithms such as AES-256
GCM~\cite{mcgrew2004security}. Here, the access key acts as encryption key, resulting
in the ciphertext $c=\mathrm{Encrypt}(n, k)$. The ciphertext is then sent to
the storage system alongside the requested storage address. The storage system
decrypts $c$ and verifies that the obtained value matches the previously sent
nonce. If successful, it returns the requested storage entry to the user.

\subsection{Ownership-based Access Control}
If a user wants to share their created assets with other well-known users,
e.g., in the context of \ac{b2b} relations, the reliance on individual access
keys that need to be shared beforehand could be impeding. An alternative approach
is to apply ownership-based verification, as described below.

It is assumed that, within certain supply chains, users access asset data
only after the asset is already delivered (and their ownership being already transferred to the
user). In this scenario, proof of asset ownership could be used for granting access
to corresponding storage entries.

\begin{figure}[t]
\centering
    \includegraphics[width=\linewidth]{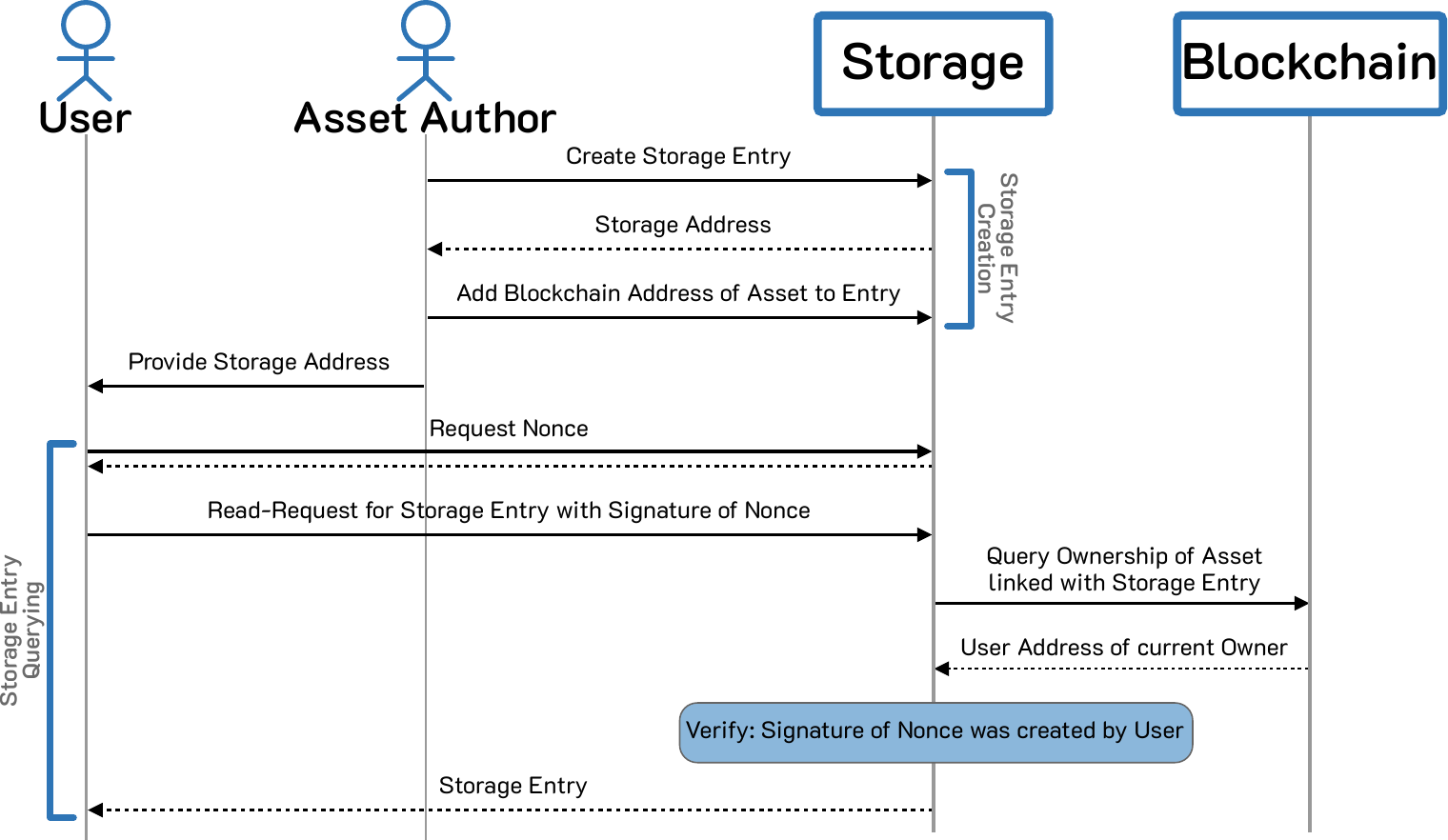}
\caption{Ownership-based authentication for storage entry access.}
\label{fig:design:ownerauth}
\end{figure}

In Hyperledger Sawtooth, the private blockchain key can
be used to sign data that is verifiable with the public blockchain address
(i.e., user address). Using this feature, storage systems can implement an
access control mechanism that verifies, if the requesting user is the current
owner of the
corresponding asset. This concept, as illustrated in
Figure~\ref{fig:design:ownerauth}, presupposes that the asset author has registered the asset's blockchain address within the storage system such that the system is capable of querying the correct asset entry.

The read request is initiated by the user with a request for a fresh nonce. The
user then signs the nonce using the private blockchain key. Thereafter, the
user submits a read request that contains the requested storage address and the
signature. The storage system queries the linked asset entry and obtains the
user address (i.e., public key) of the current owner. Using this information,
the system verifies the signature and, if successful, returns the storage
entry.

With this concept, any user that becomes an asset's owner is directly capable
of querying its storage entry. The concept can also be extended to support
ownership-based access for related assets such as a product's parent batches
and actions. Furthermore, this approach is compatible with
anonymous user addresses and hence not revealing a user's identity to the storage
system.

\section{Origin Verification}
\label{app:origin-verification}

\begin{algorithm}[t]
    \caption{Origin Trail Verification}
    \label{alg:design:origin}
    \begin{algorithmic}[3]
        \Procedure{Origin}{$a, t$}
        \Comment{Verify asset $a$ with trail $t=[]$}
        \State $p_a\gets \Call{Parent}{a}$\Comment{Find the parent batch that
        $a$ stems from.}
    \If{$p_a = Nil$}
        \If{$|t| > 0 \And \Call{IsSameProducer}{a, t_{|t|-1}} = False$}
        \State \textbf{return} $False,[]$ \Comment{Leaf asset is not created by}
        
        \Comment{root batch's producer.}
        \EndIf
        \State \textbf{return} $True, [a]\cup t$ \Comment{$a$ is the root asset.}
    \EndIf
    \If{$\Call{IntegrityCheck}{p_a} = False$}
        \State \textbf{return} $False,[]$ \Comment{Incorrect hash/signature in}
        
        \Comment{storage entry.}
    \EndIf
    \ForAll{$a'\in [a] \cup t$}\Comment{Iterate over the trail.}
        \State $c\gets \Call{FindComponent}{p_a, a'.StorageAddr}$\Comment{Check, if}
        
        \Comment{listed in parent batch.}
        \If{$c \neq Nil$}
            \If{$\Call{Hash}{a'}= c.Hash$}
                \State \textbf{return} $\Call{Origin}{p_a, [a] \cup t}$
                \Comment{All checks}
                
                \Comment{successful, jump to next parent.}
            \Else
                \State \textbf{return} $False,[]$ \Comment{Integrity violation in}
                
                \Comment{origin trail.}
            \EndIf
        \EndIf
    \EndFor
        \State \textbf{return} $False,[]$ \Comment{Parent batch neither covers asset
        $a$}
        
        \Comment{nor parts of the trail.}
\EndProcedure
    \end{algorithmic}
\end{algorithm}

For certain product types, the traversed supply chain is a factor that impacts
product originality and quality. In such cases, it is necessary to
verify an asset's origin trail.
In its most basic form, the origin trail consists of an asset's list of past
owners. While the list of user addresses from past owners can be easily obtained
from its corresponding blockchain entry, more steps must be taken for assets
that are or have been part of a batch. Such assets must consider all past
batches. Besides supply chain-specific checks for the data provided by parent
batches, it is crucial to verify that an asset's data hash is contained in all parent batches. This
verification is done by following the steps described below.

Consider an asset that is specified as a component of a batch. The batch can be
in three different states:
\begin{enumerate}
    \item The batch was directly created by a producer and, therefore,
        constitutes the origin's root.
    \item The batch is itself listed as a component of a (root) batch and not
        necessarily published in the blockchain.
    \item The batch is the result of batch splitting and, therefore, not found
        in a component list of another batch. Its blockchain entry, however, points
        to the parent batch that it was split from.
\end{enumerate}

In the case of a valid origin trail, all involved batches contain the asset's storage address and hash
value, either directly as batch component (third case),
or indirectly through a hierarchy of sub-batches that are specified in its
component list (second case).

In order to verify the integrity of the complete origin trail (from asset to
root batch), \spc~defines an algorithm that iterates
over the asset's parent batches until arriving at a root batch (cf. Algorithm~\ref{alg:design:origin}). In every
iteration, the batch's component list is checked for either the asset's storage
address and hash or for the address and hash of a parent batch from previous
iterations. If the component list holds no known entries, or if integrity
checks fail, the origin trail is considered to be corrupted. This algorithm
supports a free combination of
the static and dynamic hierarchies.
The algorithm both returns the origin trail
itself and reports, if trails are corrupted. Furthermore, the algorithm
verifies,
if the asset's producer matches the producer of the root batch.
The algorithm could be extended to produce auxiliary information, e.g., to
report the specific
batches within the trail that violate integrity checks.

\end{document}